\documentclass{emulateapj}

\begin{document}

\shortauthors{Luhman et al.}
\shorttitle{Oph 1622$-$2405: Not Planetary-Mass Binary}

\title{Oph 1622$-$2405: Not a Planetary-Mass Binary}

\author{K. L. Luhman\altaffilmark{1},
K. N. Allers\altaffilmark{2,3},
D. T. Jaffe\altaffilmark{4},
M. C. Cushing\altaffilmark{3,5,6},
K. A. Williams\altaffilmark{5,7},
C. L. Slesnick\altaffilmark{8},
and W. D. Vacca\altaffilmark{9}}

\altaffiltext{1}{Department of Astronomy and Astrophysics,
The Pennsylvania State University, University Park, PA 16802;
kluhman@astro.psu.edu.}

\altaffiltext{2}{Institute for Astronomy, The University of Hawaii at Manoa,
2680 Woodlawn Drive, Honolulu, HI 96822.}

\altaffiltext{3}{Visiting Astronomer at the Infrared Telescope Facility, which
is operated by the University of Hawaii under Cooperative Agreement no. NCC
5-538 with the National Aeronautics and Space Administration, Office of Space
Science, Planetary Astronomy Program.}

\altaffiltext{4}{Department of Astronomy, The University of Texas, Austin, TX
78712.}

\altaffiltext{5}{
Steward Observatory, The University of Arizona, Tucson, AZ 85721.}

\altaffiltext{6}{Spitzer Fellow.}

\altaffiltext{7}{Current address: Department of Astronomy, The University of 
Texas, Austin, TX 78712.}

\altaffiltext{8}{
Department of Astronomy, MS105-24, California Institute of Technology, 
Pasadena, CA 91125.}

\altaffiltext{9}{
SOFIA-USRA, NASA Ames Research Center, MS N211-3, Moffett Field, CA 94035.}

\begin{abstract}

We present an analysis of the mass and age of the young low-mass binary 
Oph~1622$-$2405.
Using resolved optical spectroscopy of the binary, we measure spectral types of 
M7.25$\pm0.25$ and M8.75$\pm0.25$ for the A and B components, respectively.
We show that our spectra are inconsistent with the spectral types of 
M9 and M9.5-L0 from Jayawardhana \& Ivanov and M9$\pm$0.5 and M9.5$\pm0.5$
from Close and coworkers. 
Based on our spectral types and the theoretical evolutionary models of
Chabrier and Baraffe, we estimate masses of $\sim0.055$ and 
$\sim0.019$~$M_\odot$ for Oph~1622$-$2405A and B, which are significantly higher
than the values of 0.013 and 0.007~$M_\odot$ derived by Jayawardhana \& Ivanov 
and above the range of masses observed for extrasolar planets 
($M\lesssim0.015$~$M_\odot$).  Planet-like mass estimates are further 
contradicted by our demonstration that Oph~1622$-$2405A is only slightly later
(by 0.5~subclass) than the composite of the young eclipsing binary brown 
dwarf 2M~0535-0546, whose components have dynamical masses of 0.034 and 
0.054~$M_\odot$.  To constrain the age of Oph~1622$-$2405, we compare the 
strengths of gravity-sensitive absorption lines  
in optical and near-infrared spectra of the primary to lines in
field dwarfs ($\tau>1$~Gyr) and members of Taurus ($\tau\sim1$~Myr) 
and Upper Scorpius ($\tau\sim5$~Myr). 
The line strengths for Oph~1622$-$2405A are inconsistent 
with membership in Ophiuchus ($\tau<1$~Myr) and instead indicate an age
similar to that of Upper Sco, which is agreement with a similar analysis
performed by Close and coworkers. We conclude that Oph~1622$-$2405 is
part of an older population in Sco-Cen, perhaps Upper Sco itself.

\end{abstract}


\section{Introduction}
\label{sec:intro}

Since the discovery of the first free-floating brown dwarfs a decade ago 
\citep{shp94,reb95,bas96}, surveys of young clusters and the field have
found brown dwarfs in increasing numbers and at decreasing masses.
A natural consequence of these growing samples of brown dwarfs has been the
identification of binary systems with progressively lower total masses
\citep[][references therein]{cha04,bur06ppv}.
The least massive binaries are potentially valuable for testing theories
for the formation of objects at the bottom of the initial mass function
\citep{bur06ppv,luh06ppv}. To produce meaningful results, 
these tests require reliable measurements of the basic properties of 
the components of substellar binaries, including spectral type, luminosity, 
age, and mass.

Oph~1622$-$2405 is one such low-mass binary that deserves careful scrutiny.
This system was discovered during a search for disk-bearing young brown 
dwarfs with the {\it Spitzer Space Telescope} \citep{all05,all06a}. 
The $1.9\arcsec$ pair was only partially resolved by
{\it Spitzer}, but was fully resolved in optical and near-infrared (IR)
images and low-resolution near-IR spectroscopy presented by 
\citet[][see also \citet{all06b}]{all05}.
Through this spectroscopy, she classified each 
component as a pre-main-sequence object and measured spectral types of 
M7.5 and M8 for the primary and secondary, respectively. 
By placing the two components on the Hertzsprung-Russell (H-R) diagram
with theoretical evolutionary models, \citet{all05} estimated masses of 0.06 
and 0.05~$M_\odot$ and a system age of $\sim40$~Myr.
In comparison, \citet{jay06a,jay06b} have reported spectral types
of M9 and M9.5-L0 and masses of 0.013 and 0.007~$M_\odot$ for the components
of Oph~1622$-$2405, leading \citet{jay06b} to characterize Oph~1622$-$2405 
as the first known planetary-mass binary. \citet{clo07} independently
discovered the binarity of Oph~1622$-$2405 and estimated masses of 
0.017 and 0.014~$M_\odot$ for its components. 

We seek to better determine the physical properties of 
the components of Oph~1622$-$2405, particularly their masses, through
new optical and near-IR spectroscopy (\S~\ref{sec:obs}). 
With the optical data, we measure the spectral types of Oph~1622$-$2405A 
and B with the optical classification scheme that has been most commonly
applied to young late-type objects and show how these types compare to
those of other young binaries through direct comparison of their optical spectra
(\S~\ref{sec:class}). We use gravity-sensitive absorption lines
in the optical and IR spectra of Oph~1622$-$2405A to constrain its age
(\S~\ref{sec:grav}). We then estimate the masses of Oph~1622$-$2405A and B
via theoretical evolutionary models and by considering the dynamical
mass measurements of the eclipsing binary brown dwarf 2MASS~J05352184-0546085 
(hereafter 2M~0535-0546, \S~\ref{sec:mass}).

\section{Observations}
\label{sec:obs}

\begin{figure}
%\epsscale{0.55}
\plotone{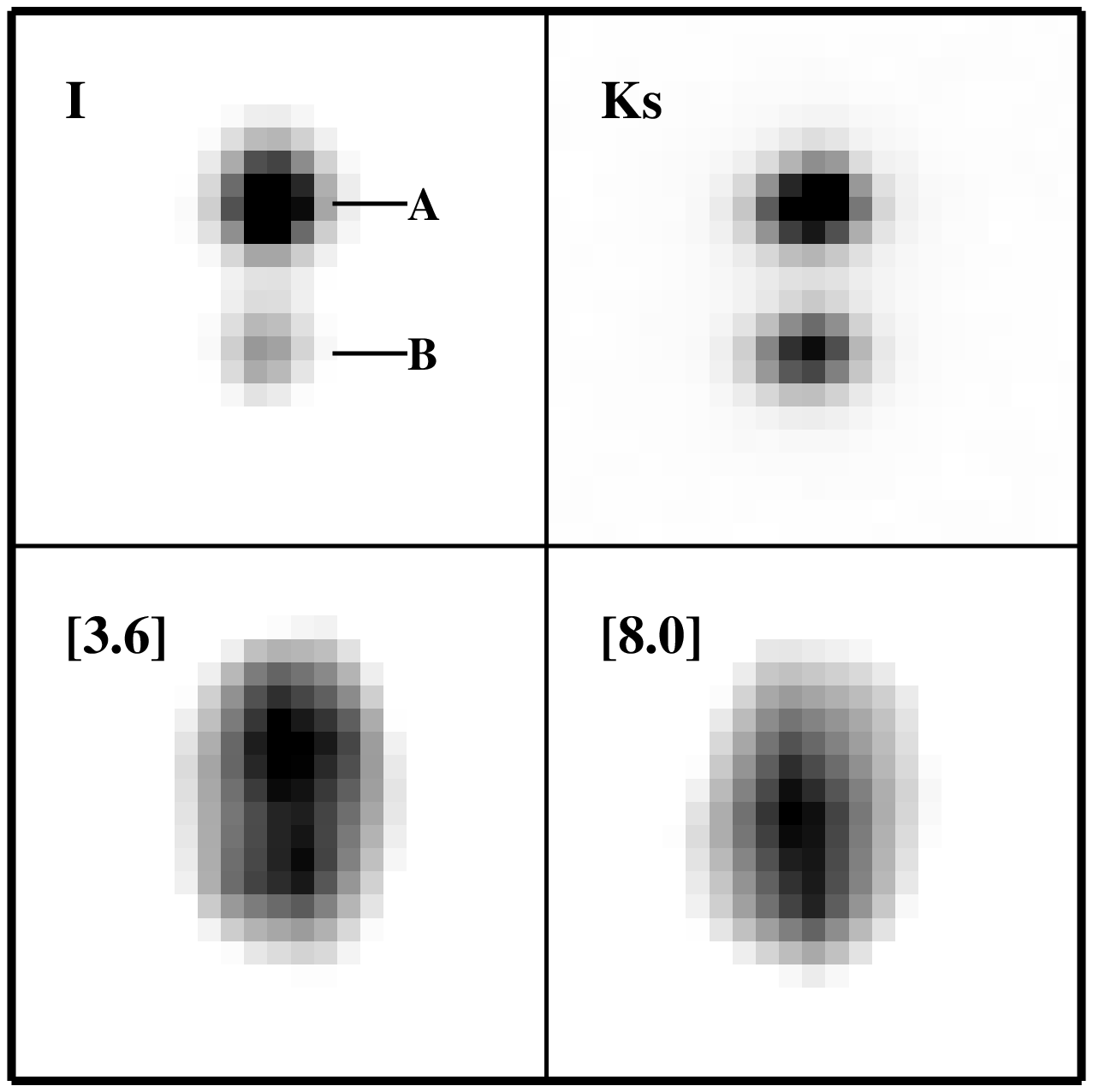}
\caption{
Discovery images of Oph~1622$-$2405A and B at $I$, $K_s$, 3.6~\micron, and 
8.0~\micron\ \citep{all05}.
The size of each image is $7\arcsec\times7\arcsec$.
North is up and east is left in these images.
}
\label{fig:image}
\end{figure}

The optical and IR images of Oph~1622$-$2405A and B (referred to as
Oph~1622$-$2405n and s, respectively, in \citet{all05})
from \citet{all05} are shown in Figure~\ref{fig:image}. 
The astrometry and photometry measured by \citet{all05} for this pair 
are provided in Table~\ref{tab:data1}. 
\citet{all06a} described the collection and reduction of the larger imaging 
survey from which these images are taken. 
\citet{clo07} measured near-IR colors for Oph~1622$-$2405A and B as well.
Their colors differed significantly from those of field M and L dwarfs, 
which they attributed to low surface gravity. However, a color difference of
that kind is not present in previous data for young late-M objects 
\citep[e.g.,][]{luh99,bri02} and the colors that we measure for
Oph~1622$-$2405A and B are similar to those of both dwarfs and 
pre-main-sequence sources.

On the nights of 2006 August 17 and 18, 
we performed optical spectroscopy on Oph~1622$-$2405A and B, respectively, with
the Low Dispersion Survey Spectrograph (LDSS-3) on the Magellan~II Telescope
and a $0.85\arcsec$ slit. This configuration resulted in a spectral resolution
of 5.8~\AA\ at 8000~\AA\ and a wavelength coverage of 5800-11000~\AA. 
For each component of the binary, we obtained one 20~min exposure with the 
slit aligned at the parallactic angle. After bias subtraction and flat-fielding,
the spectra were extracted and calibrated in wavelength with arc lamp data.
The spectra were then corrected for the sensitivity functions of the detectors,
which were measured from observations of spectrophotometric standard stars.
For comparison to Oph~1622$-$2405 in \S~\ref{sec:grav}, we will make use of a
spectrum of the eclipsing binary brown dwarf 2M~0535-0546 \citep{sta06} that 
was obtained with LDSS-3 on 2006 February 10 and spectra of GL~1111 (M6.5V)
and LHS~2243 (M8V) that were obtained with the Blue Channel spectrograph 
at the MMT on 2004 December 11 and 12, respectively.
The LDSS-3 configuration for 2M~0535-0546 was the same as for Oph~1622$-$2405.
The spectra of GL~1111 and LHS~2243 from Blue Channel have a resolution 
of 2.6~\AA\ at 8000~\AA\ and a wavelength coverage of 6300-8900~\AA.

We obtained near-IR spectra of Oph~1622$-$2405A and B with the spectrometer 
SpeX \citep{ray03} at the NASA Infrared Telescope Facility (IRTF) on the
nights of 2006 June 22 and 24.
The instrument was operated in the SXD mode with a $0.5\arcsec$ slit,
producing a wavelength coverage of 0.8-2.5~\micron\ and a resolving power of
$R=1200$. With the slit rotated to encompass both components of the pair,
we obtained 10 2-minute exposures during sequences of dithers 
between two positions on the slit on each night.
For comparison purposes, we also make use of a spectrum of the Taurus
member CFHT~4 that was obtained on the night of 2006 January 4 with the
same instrument configuration as used for Oph~1622$-$2405, except with a
$0.3\arcsec$ slit ($R=2000$). These SpeX data were reduced with the Spextool 
package \citep{cus04} and corrected for telluric absorption \citep{vac03}.

\section{Analysis}
\label{sec:analysis}

\subsection{Spectral Classification}
\label{sec:class}

\begin{figure}
%\epsscale{0.6}
\plotone{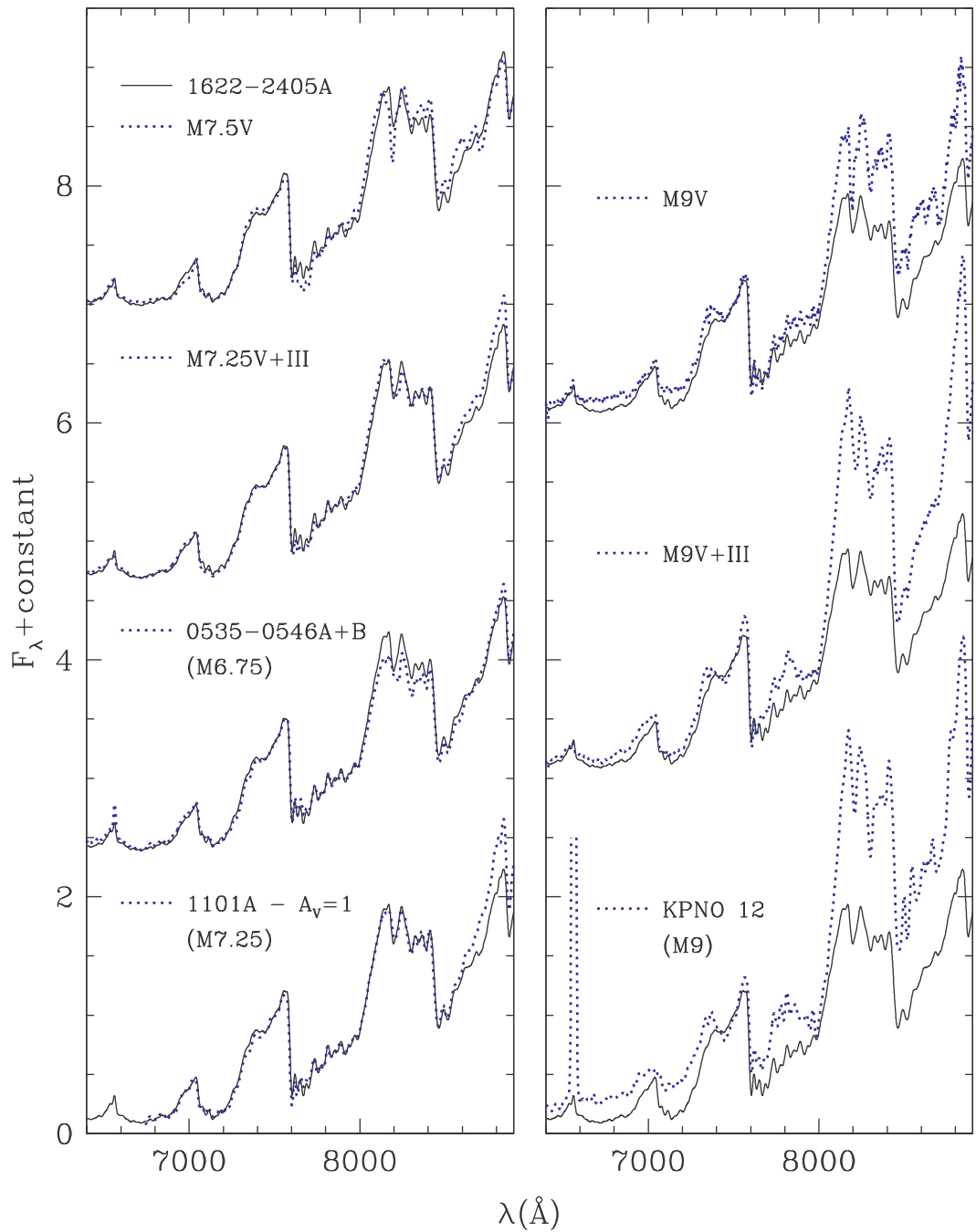}
\caption{
Optical spectrum of Oph~1622$-$2405A ({\it solid lines}) and seven comparison 
spectra ({\it dotted lines}).
If dwarf standards are used to classify this object, then M7.5V provides the 
best match. When we instead use averages of dwarfs and giants as standards,
we derive a spectral type of M7.25.
Oph~1622$-$2405A is slightly later than the composite spectrum of the 
eclipsing binary 2M~0535-0546, whose components have dynamical masses of 
0.034 and 0.054~$M_\odot$ \citep{sta06}. 
The spectrum of Oph~1622$-$2405A agrees well with the primary 
in the young binary 2M~1101-7732 \citep[M7.25,][]{luh04bin}. 
Although \citet{jay06a} and \citet{clo07} each reported a spectral type of M9 
for Oph~1622$-$2405A, its spectrum differs significantly from those of 
M9V, M9V+M9III, and the M9 Taurus member KPNO~12 \citep{luh03tau}. 
The data are displayed at a resolution of 18~\AA\ and are normalized at 
7500~\AA.
}
\label{fig:op1}
\end{figure}

\begin{figure}
%\epsscale{0.6}
\plotone{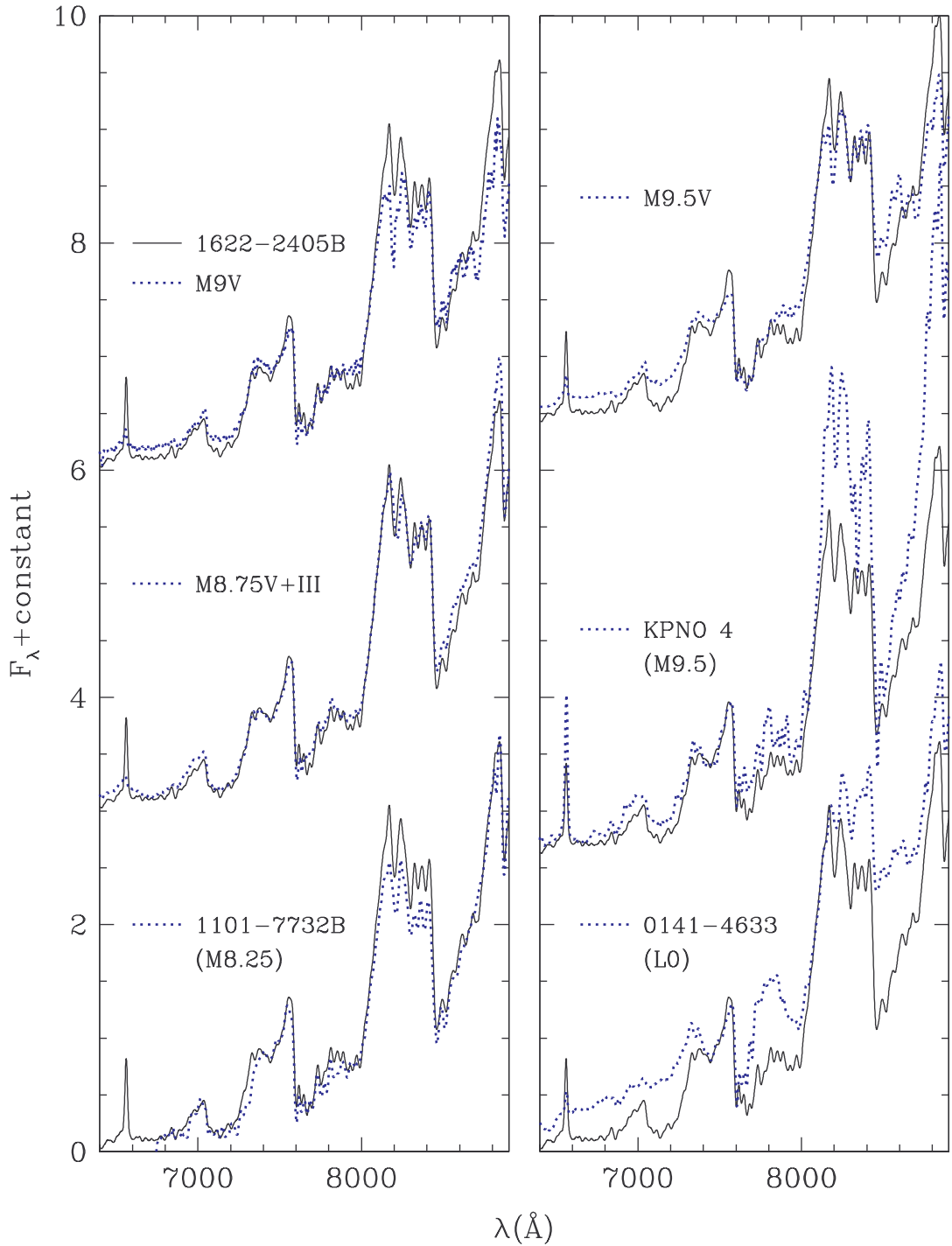}
\caption{ 
Optical spectrum of Oph~1622$-$2405B ({\it solid lines}) and six comparison 
spectra ({\it dotted lines}).
If dwarf standards are used to classify Oph~1622$-$2405B, M9V provides the
best match. When we instead use averages of dwarfs and giants as standards, 
we derive a spectral type of M8.75.
The average of the dwarf and giant agrees better with the spectrum of 
Oph~1622$-$2405B than the dwarf, as expected for a pre-main-sequence object
of this kind \citep{luh99}. The spectrum of Oph~1622$-$2405B is later than the
secondary in the young binary 2M~1101-7732 \citep[M8.25,][]{luh04bin}.
Although \citet{jay06b} and \citet{clo07} reported spectral types of M9.5-L0 
and M9.5$\pm$0.5 for Oph~1622$-$2405B, respectively, its spectrum differs 
significantly from those of M9.5V, the M9.5 Taurus member KPNO~4 
\citep{bri02}, and the young L0 object 2M~0141-4633 \citep{kir06}. The data 
are displayed at a resolution of 18~\AA\ and are normalized at 7500~\AA.
}
\label{fig:op2}
\end{figure}

Spectral types of late-M dwarfs and giants are defined at red optical 
wavelengths \citep{kir91}. Averages of optical spectra of standard dwarfs and 
giants agree well with data for late-M pre-main-sequence objects 
\citep{luh97,luh98,luh99} and are the basis 
of most of the published optical spectral types for low-mass members of nearby
star-forming regions \citep[e.g.,][]{bri02,luh03ic,luh04cha,luh06tau}.
We have applied this classification scheme to our optical 
spectra of Oph~1622$-$2405A and B, arriving at spectral types of M7.25$\pm0.25$
and M8.75$\pm0.25$, respectively. 
If we use dwarfs alone to classify our spectra as done for this binary
by \citet{jay06b}, then we derive nearly the same spectral types, namely
M7.5 and M9. Figs.~\ref{fig:op1} and \ref{fig:op2} show comparisons of 
Oph~1622$-$2405A and B to these best-matching standards. The averages of dwarfs
and giants match the target spectra more closely than dwarfs alone, 
in agreement with previous work \citep[e.g.,][]{luh99}.  Thus, we adopt the 
types based on the comparisons to averages of dwarf and giant standards.
By doing so, our classifications can be reliably compared to types of
most known young late-type objects, which have been measured in the same way.

In Figure~\ref{fig:op1}, we compare the spectrum of Oph~1622$-$2405A to data 
for the composite of the young eclipsing binary 2M~0535-0546 (\S~\ref{sec:obs})
and the primary in the young wide binary 2M~1101-7732 \citep{luh04bin}. 
This comparison demonstrates that Oph~1622$-$2405A is slightly later than 
2M~0535-0546A+B (which we classify as M6.75) and has the same spectral type as 
2M~1101-7732A \citep[M7.25,][]{luh04bin}. 
Meanwhile, Oph~1622$-$2405B is slightly later than 2M~1101-7732B 
\citep[M8.25,][]{luh04bin} based on the comparison of their spectra in 
Figure~\ref{fig:op2}.

Our optical spectral types of M7.25$\pm0.25$ and M8.75$\pm0.25$ 
for Oph~1622$-$2405A and B are significantly earlier than the optical types 
of M9 and M9.5-L0 from \citet{jay06a,jay06b}. 
This is true even if we use dwarf standards as done in those studies. 
As shown in Figure~\ref{fig:op1}, the spectrum of Oph~1622$-$2405A 
is poorly matched by both M9V and M9V+M9III, as well as
the M9 Taurus member KPNO~12 \citep{luh03tau}.
Similarly, the spectrum of Oph~1622$-$2405B is earlier than 
KPNO~4 (Figure~\ref{fig:op2}), which is the prototypical 
pre-main-sequence representative of the M9.5 spectral class \citep{bri02}.
Oph~1622$-$2405B also differs significantly from M9.5V
and the young L0 object 2MASS~01415823-4633574 
\citep[hereafter 2M~0141-4633,][]{kir06}. For instance, a defining
characteristic of the transition from M to L types for dwarfs is the
disappearance of TiO absorption at 7000-7200~\AA\ \citep{kir99}, 
and yet the TiO in Oph~1622$-$2405B is strong, indicating that 
a dwarf-based spectral type of M9.5-L0 is not appropriate.
Our results for Oph~1622$-$2405~A and B are similar to those of \citet{all07}
for another object classified by \citet{jay06a}, 2M~1541-3345, which is
a disk-bearing source in the vicinity of the Lupus clouds \citep{all05,all06a}. 
\citet{jay06a} reported a spectral type of M8 for this object while
\citet{all07} classified it is M5.75$\pm$0.25 with spectra and methods
like those used in this work. 

We now compare our optical classifications of Oph~1622$-$2405A and B
to previous near-IR observations.
Our optical types of M7.25$\pm0.25$ and M8.75$\pm0.25$ for Oph~1622$-$2405A and
B are consistent with the IR types of M7.5$\pm1$ and M8$\pm1$ from 
\citet{all05} and M7$\pm1$ and M8$\pm1$ from \citet{all06b}, which were 
measured by comparison to young objects that have been optically 
classified with the same methods employed in this work. 
\citet{bra06} and \citet{clo07} also presented near-IR spectra for 
Oph~1622$-$2405A and B. \citet{bra06} did not compare their spectra 
to data for classification standards and thus did not measure spectral types. 
\citet{clo07} found that Oph~1622$-$2405B exhibited similar IR spectral
features as the young L0 source 2M~J0141-4633, and thus classified the former
as M9.5$\pm0.5$. Because the spectral features of the primary indicated
a slightly earlier type, they classified it as M9$\pm0.5$. However, given that
\citet{clo07} did not present a comparison of Oph~1622$-$2405 to earlier types, 
it is unclear if earlier types would have provided better or 
worse matches to their data. In fact, as shown in Figure~\ref{fig:op2},
our optical spectrum of Oph~1622$-$2405B is very different from the spectrum
of 2M~J0141-4633 from \citet{kir06}, demonstrating that they do not have the
same optical spectral types. 

\citet{bra06} and \citet{clo07} estimated 
effective temperatures and surface gravities by comparing their data
to synthetic spectra. However, because of known deficiencies in theoretical 
spectra of late-type objects \citep{leg01}, temperature and gravity estimates 
of this kind are subject to systematic errors, and thus the accuracies of these
estimates are unknown. In addition, the temperature and gravity estimates
from those studies cannot be reliably compared to those of other young
late-type objects unless the latter are derived with the same 
spectral features, model spectra, and fitting procedures. 

\subsection{Age}
\label{sec:grav}

\citet{all05} and \citet{all06b} determined that the components of
Oph~1622$-$2405 are pre-main-sequence objects rather than field dwarfs by
performing low-resolution near-IR spectroscopy and detecting the presence 
of triangular $H$-band continua \citep{luc01}. 
This spectral characteristic is present during most of the pre-main-sequence
evolution of late-type objects, thus constraining the age of Oph~1622$-$2405 
to be $\tau\lesssim100$~Myr \citep{kir06,all06b}.
To further constrain the age of this system, we examine additional
gravity-sensitive absorption lines in our higher-resolution optical and near-IR 
spectra, namely K~I, Na~I, and FeH \citep{mar96,luh98,gor03,mc04}. 

In this analysis, we consider only the primary because 
its data exhibit better signal-to-noise.
For comparison to Oph~1622$-$2405A, we select representatives of three
distinct luminosity classes and ages: members of the Taurus star-forming
region \citep[$\tau\sim1$~Myr,][]{bri02,luh03tau}, members of the 
Upper Scorpius OB association \citep[$\tau\sim5$~Myr,][]{pm06},
and field dwarfs ($\tau>1$~Gyr). We require that these objects have spectral
types that are within 0.5 subclass of the spectral type of the primary
so that our comparison is sensitive to variations in surface gravity alone.
For Taurus, we use the optical spectrum of 2MASS~04484189+1703374 
\citep[M7,][]{luh06tau}
from \citet{luh06tau} and our IR spectrum of CFHT~4 \citep[M7,][]{bri02}.
Field dwarfs are represented by an average of our optical spectra of
GL~1111 \citep[M6.5V,][]{hen94} and LHS~2243 \citep[M8V,][]{kir95} 
and the IR spectrum of vB~8 \citep[M7,][]{kir91} from \citet{cus05}. 
For Upper Sco, we use the $J$-band spectrum of U~Sco~CTIO~128
\citep[M7,][]{ard00} from \citet{sle04}.
To enable a reliable comparison of absorption line strengths, spectra for
a given wavelength range have been smoothed to a common spectral resolution. 
Although optical spectra for Upper Sco are available from \citet{sle06},
we do not include them in this comparison because they have significantly 
lower spectra resolution that the other optical data we are examining,
and we prefer to make these comparisons at the highest possible resolution.

\begin{figure}
%\epsscale{0.8}
\plotone{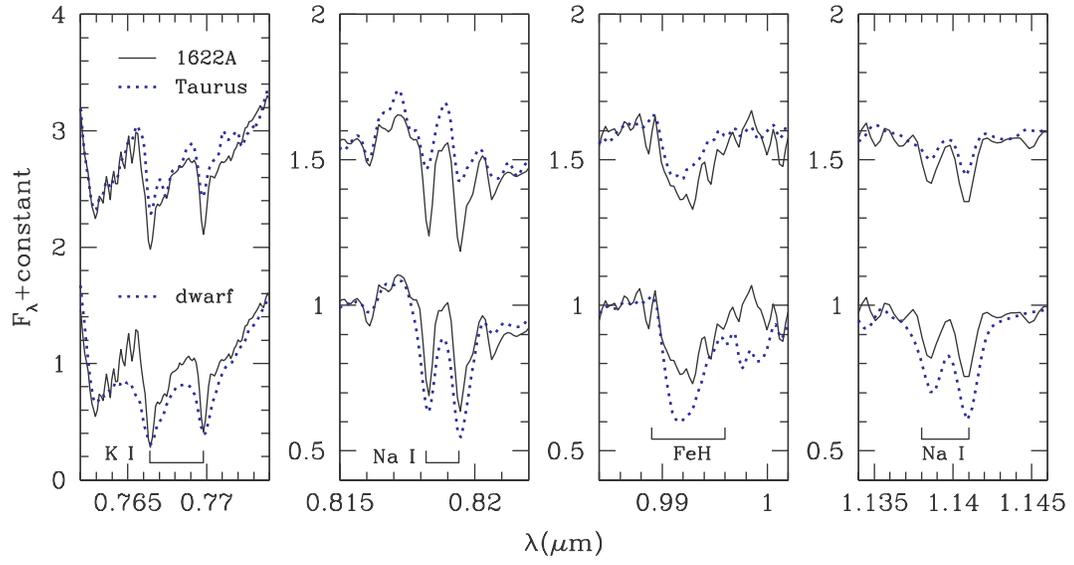}
\caption{ 
Gravity-sensitive absorption lines for Oph~1622$-$2405A ({\it solid lines}),
a Taurus member ($\tau\sim1$~Myr, {\it upper dotted lines}), and a
field dwarf ($\tau>1$~Gyr, {\it lower dotted lines}).
}
\label{fig:lines1}
\end{figure}

\begin{figure}
%\epsscale{0.55}
%\vspace*{-10mm}
\plotone{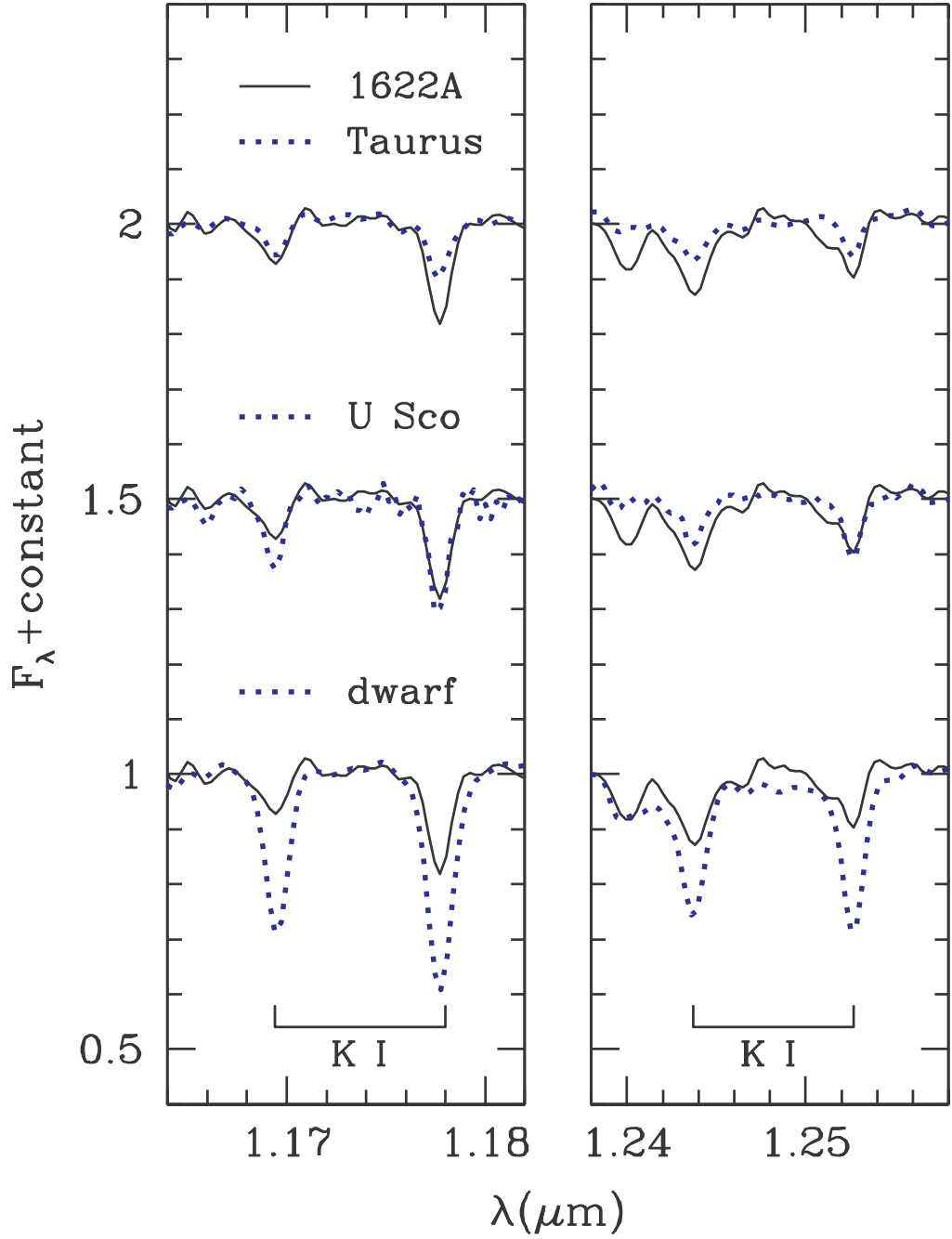}
\caption{ 
Gravity-sensitive absorption lines for Oph~1622$-$2405A ({\it solid lines}),
a Taurus member ($\tau\sim1$~Myr, {\it upper dotted lines}), 
an Upper~Sco member ($\tau\sim5$~Myr, {\it middle dotted lines}), 
and a field dwarf ($\tau>1$~Gyr, {\it lower dotted lines}).
The data in this diagram and in Figure~\ref{fig:lines1} indicate that 
Oph~1622$-$2405A is a pre-main-sequence source with an age of $\sim5$~Myr.
}
\label{fig:lines2}
\end{figure}

In Figs.~\ref{fig:lines1} and \ref{fig:lines2}, 
we compare the spectra for Taurus, Upper Sco, and field dwarfs to
the spectrum of Oph~1622$-$2405A for wavelength ranges encompassing 
K~I, Na~I, and FeH. For all of these transitions, Oph~1622$-$2405A 
exhibits stronger lines than the Taurus members and weaker lines than
the field dwarfs, indicating that it is above the main sequence 
($\tau\lesssim100$~Myr) but older than Taurus ($\tau>1$~Myr). 
For the subset of comparisons in which Upper Sco is represented, the line
strengths are similar between Oph~1622$-$2405A and the Upper Sco member.
If we degrade the optical spectrum of Oph~1622$-$2405A to the lower
spectral resolution of optical data in Upper Sco from \citet{sle06}, we
also find similar line strengths for the optical transitions of Na~I and K~I.
Thus, the gravity-sensitive lines in the spectra of Oph~1622$-$2405A 
suggest an age similar to that of Upper Sco ($\tau\sim5$~Myr) with 
an upper limit that is undetermined, but probably no more than a few tens 
of millions of years. 

Using their near-IR spectra of Oph~1622$-$2405A and B, 
\citet{bra06} compared the equivalent widths of gravity-sensitive lines 
between Oph~1622$-$2405 and field dwarfs, demonstrating that the binary's
components have lower gravities and hence younger ages than dwarfs,
which is consistent with the results in this work, \citet{all05}, 
\citet{all06b}, and \citet{clo07}.
However, their analysis did not further refine the age of the system and 
they did not claim to accurately measure the gravities and ages of
Oph~1622$-$2405A and B, and instead assumed membership in Ophiuchus.
On the other hand, \citet{clo07} constrained the gravity of Oph~1622$-$2405A
and B more tightly by comparing their data to KPNO~4 ($\tau\sim1$~Myr) 
and $\sigma$~Ori~51 ($\tau\sim5$~Myr). This comparison indicated that 
Oph~1622$-$2405A and B are older than KPNO~4 and similar in age to 
$\sigma$~Ori~51, which is in agreement with the age constraints that we
have derived in this section.

\subsection{Membership}

The analysis of gravity-sensitive lines in \S~\ref{sec:grav}
demonstrated that Oph~1622$-$2405A is older than Taurus ($\tau\sim1$~Myr), 
which in turn is older than the stellar population within 
the Ophiuchus cloud core \citep[$\tau<1$~Myr,][]{lr99}. 
Thus, Oph~1622$-$2405 is not a member of the current generation of stars
forming within the Ophiuchus cloud, which is consistent with the
low extinction of this binary ($A_V<1$) and its large angular 
distance from the cloud core ($\theta\sim0\fdg5$). 
Instead, Oph~1622$-$2405 probably is part of an older population of stars 
in the Sco-Cen complex, which contains several neighboring and overlapping 
generations of stars with ages from $<1$ to 20~Myr \citep{pm06}. 
For instance, Oph~1622$-$2405 is within the area encompassed by 
known members of Upper Sco \citep{sle06,pm06} and is near a population of 
exposed young stars distributed across the front of the Ophiuchus cloud, 
which is coeval with Upper Sco and may be an extension of it \citep{wil05}.

Although we cannot definitively identify the origin of Oph~1622$-$2405 nor
measure its distance, membership in Upper Sco is likely based on its
location and surface gravity diagnostics. Indeed, this evidence of membership
in Upper Sco is the same as for previously reported
late-type members \citep{mar04,sle06}. Therefore, for the purpose of
estimating their luminosities, we assign to Oph~1622$-$2405A and B the distance
of Upper Sco, which extends from 125 to 165~pc \citep{pm06}. 
Based on the comparison of the optical spectra of Oph~1622$-$2405A and B to 
spectra of other young late-type objects in \S~\ref{sec:class}, we find
that the extinction of each component is $A_V<1$. Therefore, 
we adopt $A_V=0.5\pm0.5$ for measuring their luminosities. 
The remaining details of the luminosity estimates are provided by 
\citet{all06a}. Our luminosity measurements for Oph~1622$-$2405A and B
are listed in Table~\ref{tab:data2}. 

\begin{figure}
%\epsscale{0.55}
\plotone{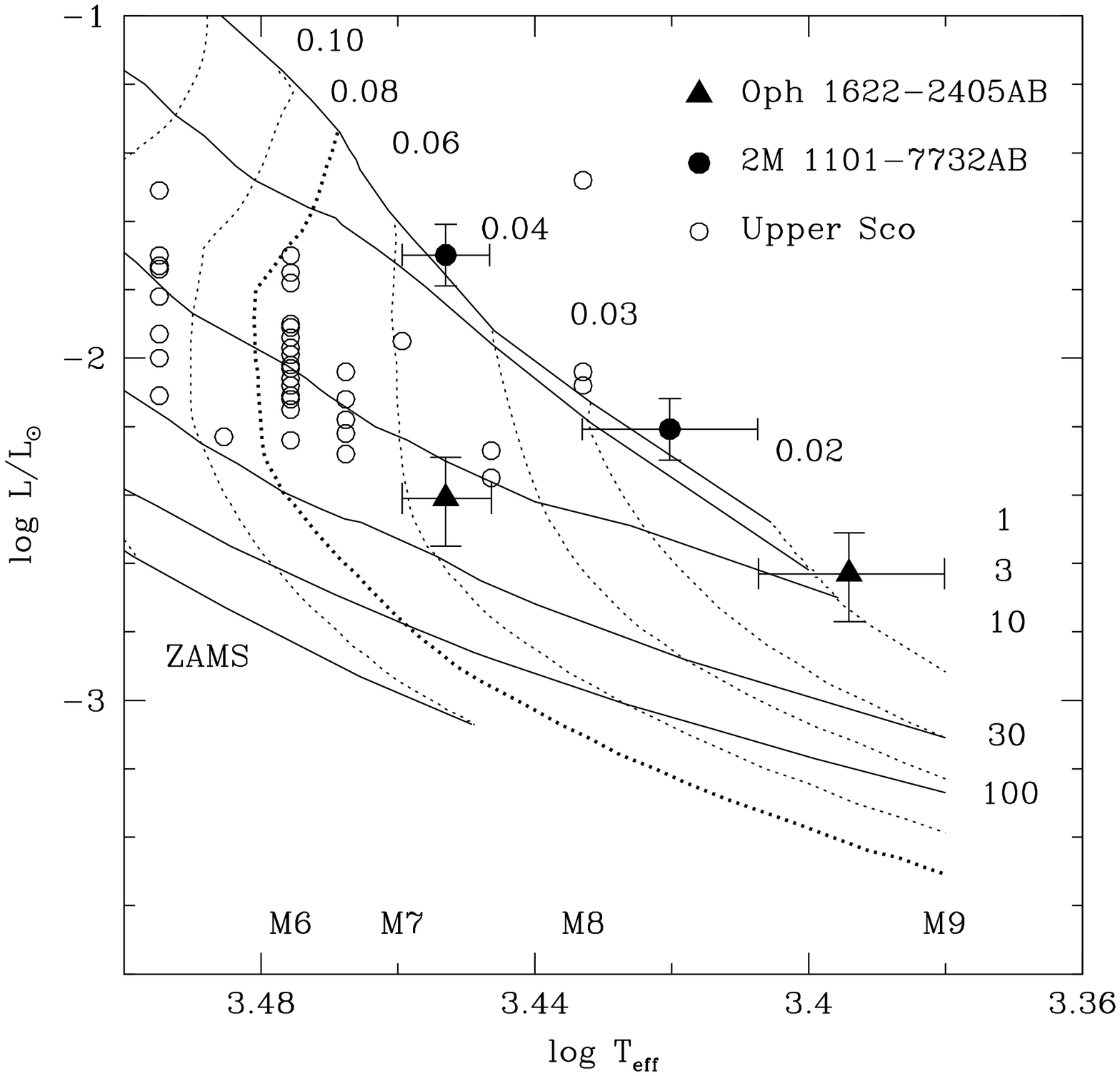}
\caption{ 
H-R diagram for Oph~1622$-$2405A and B 
\citep[{\it filled circle},][Table~\ref{tab:data2}]{all05}, 
2M~1101-7732A and B \citep[{\it triangle},][]{luh04bin}, 
and members of Upper Sco \citep[{\it circles},][]{sle06}
shown with the theoretical evolutionary models of
\citet{bar98} ($M/M_\odot>0.1$) and \citet{cha00} ($M/M_\odot\leq0.1$),
where the mass tracks ({\it dotted lines}) and isochrones ({\it solid lines})
are labeled in units of $M_\odot$ and Myr, respectively.
According to this diagram, Oph~1622$-$2405A and B have masses that
are similar to those of 2M~1101-7732A and B. In addition, the 
positions of Oph~1622$-$2405A and B are consistent with an extension of the
Upper Sco sequence to lower masses.
}
\label{fig:hr}
\end{figure}

To examine the ages implied by their luminosities, we plot Oph~1622$-$2405A 
and B on the H-R diagram in Figure~\ref{fig:hr} with the evolutionary models of 
\citet{bar98} and \citet{cha00}. We have converted our optical spectral types 
to effective temperatures with a temperature scale that is compatible
with these models for young objects \citep{luh03ic}.
The data and models in Figure~\ref{fig:hr} imply ages of 10-30~Myr for
the primary and 1-20~Myr for the secondary. Thus, these results are 
consistent with coevality for the two objects, which is expected for a
binary system. These ages are somewhat older than the canonical value of 
5~Myr that is usually quoted for Upper Sco, but the age of a young population is
sensitive to how it is defined, the mass range of objects considered, and
the choice of models. To reliably compare the inferred ages of
Oph~1622$-$2405A and B to those of Upper Sco members, the luminosities and 
temperatures of this binary should be compared directly to those of late-type
members of Upper Sco. We do this by including in Figure~\ref{fig:hr}
the late-type members of Upper Sco from \citet{sle06}. The lower limits
of the sequence in temperature and luminosity are reflections of the 
detection limits of the survey from \citet{sle06}. An extension of the 
sequence below these limits in a manner that is parallel to the theoretical
isochrones would encompass both components of Oph~1622$-$2405. In other words,
the model ages of the more massive members of the Upper Sco sequence from 
\citet{sle06} are consistent with the model ages of Oph~1622$-$2405A and B.
Similarly, the binary components fall within the sequence of low-mass
Upper Sco members in the color-magnitude diagram from \citet{mar04}.

\subsection{Mass}
\label{sec:mass}

In addition to ages, the evolutionary models in Figure~\ref{fig:hr}
also provide estimates of masses, implying values of 0.055$\pm$0.01 and 
$0.019^{+0.01}_{-0.005}$~$M_\odot$ for Oph~1622$-$2405A and B, respectively. 
The quoted uncertainties reflect only the uncertainties in spectral types
and luminosities. Additional systematic errors could be introduced by the
adopted temperature scale and evolutionary models. 
However, the sizes of these systematic errors are probably not large, as 
demonstrated by various observational tests \citep{luh03ic,luh06abdor}.
The mass estimated for Oph~1622$-$2405A by \citet{all05} is similar to our
value, while her estimate for the secondary was twice our value because of
the earlier spectral type that she derived. 
Meanwhile, our estimates for Oph~1622$-$2405A and B are significantly higher
than the masses of 0.013 and 0.007~$M_\odot$ from \citet{jay06b}. 
Our estimate for the primary is also much higher than the value of 
0.017~$M_\odot$ from \citet{clo07}, while our mass for the secondary is
only slightly higher than their mass of 0.014~$M_\odot$.
The validity of higher estimates is supported by the fact that
Oph~1622$-$2405A is only slightly cooler than the composite of the eclipsing
binary brown dwarf 2M~0535-0546A+B (Figure~\ref{fig:op1}) and thus 
should have a comparable mass \citep[$M=0.054$ and 0.034~$M_\odot$,][]{sta06}.

The spectral types, temperatures, luminosities, and masses of Oph~1622$-$2405A 
and B produced by our analysis are compiled in Table~\ref{tab:data2}.

\section{Conclusions}

Using optical spectroscopy, we have measured spectral types for 
the young binary Oph~1622$-$2405 that are significantly earlier than those
reported by \citet{jay06a,jay06b} and \citet{clo07}. 
As a result, our mass estimates for these objects ($M=0.055$ and 
0.019~$M_\odot$) are higher than those from 
\citet[][$M=0.013$ and 0.007~$M_\odot$]{jay06a,jay06b} 
and \citet[][$M=0.017$ and 0.014~$M_\odot$]{clo07} and are above the range of
planetary masses \citep[$M\lesssim0.015$~$M_{\odot}$,][]{mar05}. 
Through a direct comparison of their spectra, we find that the primaries
in Oph~1622$-$2405 and the young wide binary 2M~1101-7732 have the same spectral
types while Oph~1622$-$2405B is only slightly later than 2M~1101-7732B,
which strongly indicates that these two binaries have similar masses. 
Our analysis of gravity-sensitive absorption lines in the spectra of 
Oph~1622$-$2405A have demonstrated that this system is too old to be a member
of the Ophiuchus star-forming region ($\tau<1$~Myr). 
Instead, the age constraints from those data combined with the position of 
Oph~1622$-$2405 on the H-R diagram are consistent with membership in Upper Sco 
($\tau\sim5$~Myr)\footnote{Using the same spectral classification methods
shown in this work, \citet{all07} concluded that another object from 
\citet{jay06a}, 2M~1541-3345, is also earlier, older, and more massive than 
reported in that study.}. Additional observations (e.g., proper motions, 
radial velocities) are
needed to better determine the origin and membership of this binary system.
If the distance of Upper Sco is adopted for Oph~1622$-$2405, then the 
separation of $1.9\arcsec$ for this binary corresponds to $\sim300$~AU, making
it the second young wide binary brown dwarf to be found.
Thus, Oph~1622$-$2405 is very similar to 2M~1101-7732 in both mass and 
separation but is somewhat older (5~Myr versus 1~Myr). 
Given the advanced age of this system compared to most disk-bearing stars
and brown dwarfs, the disk detected in Oph~1622$-$2405 by \citet{all06a}
is a valuable laboratory for studying the evolution of brown dwarf disks.

\acknowledgements
We thank Davy Kirkpatrick for providing his spectrum of 2M~J0141-4633
and Laird Close for providing his results prior to publication.
K. L. was supported by grant AST-0544588 from the National Science Foundation.
We thank Ivelive Momcheva for providing the telescope time for the LDSS-3
observations of Oph~1622$-$2405.
This work is supported by NASA through the Spitzer Space Telescope Fellowship
Program, through a contract issued by the Jet Propulsion Laboratory, California
Institute of Technology under a contract with NASA.

\clearpage
\begin{deluxetable}{llllllll}
\tabletypesize{\scriptsize}
\tablewidth{0pt}
\tablecaption{Astrometry and Photometry for Oph 1622$-$2405\label{tab:data1}}
\tablehead{
\colhead{Component} &
\colhead{$\alpha$(J2000)} &
\colhead{$\delta$(J2000)} &
\colhead{$I$} &
\colhead{$J$} &
\colhead{$H$} &
\colhead{$K_s$} &
\colhead{[3.6]}}
\startdata
A & 16 22 25.2 & -24 05 13.7 & 17.78$\pm$0.10 & 14.53$\pm$0.03 & 14.01$\pm$0.03 & 13.55$\pm$0.03 & 13.02$\pm$0.11 \\
B & 16 22 25.2 & -24 05 15.6 & 18.98$\pm$0.10 & 15.24$\pm$0.03 & 14.64$\pm$0.03 & 14.03$\pm$0.03 & 13.22$\pm$0.11 \\
\enddata
\end{deluxetable}

\begin{deluxetable}{lllll}
\tabletypesize{\scriptsize}
\tablewidth{0pt}
\tablecaption{Properties of Oph 1622$-$2405\label{tab:data2}}
\tablehead{
\colhead{} &
\colhead{Spectral} &
\colhead{$T_{\rm eff}$\tablenotemark{a}} &
\colhead{} &
\colhead{Mass} \\
\colhead{Component} &
\colhead{Type} &
\colhead{(K)} &
\colhead{log $L/L_\odot$\tablenotemark{b}} &
\colhead{($M_\odot$)}}
\startdata
A & M7.25$\pm$0.25 & 2838 & -2.41$\pm$0.13 & 0.055$\pm$0.01 \\
B & M8.75$\pm$0.25 & 2478 & -2.63$\pm$0.13 & $0.019^{+0.01}_{-0.005}$ \\
\enddata
\tablenotetext{a}{Temperature scale from \citet{luh03ic}.}
\tablenotetext{b}{Based on an assumed distance of 145$\pm$20~pc.}
\end{deluxetable}
\clearpage

\end{document}